# Fuzzy Logic and Quantum Measurement Formulation

N. Abbasvandi[1], M. J. Soleimani[2], Shahidan Radiman[1]

[1]*School of Applied Physics, Universiti Kebangsaan Malaysia*
*43600 UKM, Bangi Selangor, Malaysia*
*E-mail address: niloofar@siswa.ukm.edu.my*
[2]*Department of Physics, University of Malaya*
*50602, KL, Malaysia*

The Von Neumann quantum measurement theory and Zurek reformulation are based on an assumption that the quantum system, apparatus and environment obey the quantum mechanics rules. According to the Zurek theory the observers typically interact with their surrounding environments. In this article, we give a more realistic picture of the quantum measurement theory; we have proposed an improvement to Zurek quantum measurement theory based on the fuzzy logic and fuzzy set theory.

## 1. Introduction

Quantum mechanics plays a fundamental role in physics for describing the universe. It goes back to more than two centuries ago when a wave theory of light was proposed by Hooke, Huygens and Euler [1]. Quantum mechanics foundations started with Max Planck who based his attention on problem of black body radiation in 1900 [2,3], and was interpreted realistically by Einstein in 1905 [4].

Quantum mechanics was developed by Bohr, Heisenberg, de Broglie, Schrödinger, Born, Dirac, Hilbert, Sommerfeld, Dyson, Wien, Pauli, Von Neumann and others during the first three decades of the 20$^{th}$ century [5, 6, 7, 8, 9, 10]. Review of the theory development [11, 12, 13, 14, 15, 16, 17], shows us the successful story, but we cannot ignore that some questions remained unsolved regarding the fundamental features of this theory [18, 19, 20, 21]. One of the most important of them is the so-called measurement problem; has been a source of endless speculation. In fact, the measurement problem has been raised based on the superposition principle, namely: *A quantum system exists in all its particular possible states simultaneously and a result corresponding to only one of the possible configurations will be given when the system is measured or observed* [13]. This is the wave function reduction concept (equivalent terms to collapse and contraction of the wave function are also used) which has been introduced by Heisenberg in 1927 [22] and formulated by Von Neumann in 1932 [23]. Von Neumann postulated that the wave function of the system under consideration evolves as a function of time in two ways:

1) In a regular manner, i.e. in accordance with the Schrödinger equation
2) In a singular manner, that cannot be described by the standard quantum theory equations [24].

The Von Neumann model of quantum measurement was developed based on the above postulates which have provided the standard setting for the exploration of the role of observers [23]. In fact, the measurement problem has been focused on discussions of the interpretation of quantum theory since the 1920's in standard form, but the concept leads to some conceptual difficulties.

In this regard, during the past two decades, Zurek developed a new theory of measurement for removing the non-clarified aspects of the Von Neumann measurement theory [29]. The key idea promoted by him, is the insight that the realistic quantum systems are not only ever isolated, but also they interact continuously with the surrounding environment [42]. The approach by Zurek leads us to the concept of decoherence [25]. Although, other physicists try to address the whole conceptual questions of the measurement problem using decoherence concept, some conceptual problems still need clarification, reformulation and reinterpretation. In this manuscript we reformulate and reinterpret the Von Neumann measurement theory, considering conceptual picture of Zurek measurement theory by using a mathematical tool which is known as Fuzzy sets theory based on the Fuzzy logic.

The paper is organized as follows: In section 2, we review Von Neumann quantum measurement model. In section 3 we cover Zurek reformulation and decoherence. Section 4 deals with Fuzzy logic and Fuzzy set theory. We develop Fuzzy quantum measurement in section 5. The last part is the discussion and conclusion.

## 2. Quantum Measurement – Von Neumann Model

The quantum theory was developed during the first half of the 20[th] century. The conventional quantum mechanics is based on several postulates [26] like classical mechanics which are as follows:

1. To every state of a physical system there is a function $|\Psi\rangle$ associated to and defining the state.

2. If $H_1$ is the Hilbert space associated with physical system $S_1$ and $H_2$ is the Hilbert space corresponding to the other physical system $S_2$, then the composite system $S_1 + S_2$ will be associated with the tensor product of the two Hilbert vector spaces $H_1 \otimes H_2$.

3. To every observable of a physical system is associated a self- ad joint (or Hermitian) operator allowing a compatible set of eigenfunctions.

4. The time evolution of a quantum state is governed by a unitary linear transformation operator $\hat{U}$. if $\psi(t)$ is the probability amplitude of a quantum state at time t, then $\psi(t + \Delta t)$ is its probability amplitude at a later time $t + \Delta t$, so that $|\psi(t + \Delta t)\rangle = \hat{U}|\psi(t)\rangle$

5. As a result of a measuring process performed upon an observable $a$, we will obtain only the eigenvalues of the Hermitian operator $\hat{A}$, associated to the observable. The probability of getting an Eigenvalue $a_k$ corresponding to the discrete spectrum is $|C_k|^2$ and the probability of getting an Eigenvalue $a_k$ corresponding to the continuous spectrum within an interval $d\alpha$ is $|C_\alpha|^2 d\alpha$

We consider $H$ as Hilbert space and $\{|a_k\rangle\}_{k=1}^{\infty}$ to be a sequence of elements of $H$ and we consider a system to be measured. The state of the system is defined as $|\Psi\rangle$ which is a vector in Hilbert space. If the set of vectors $\{|a_k\rangle\}_{k=1}^{\infty}$ is an orthonormal basis for this space, one can express $|\Psi\rangle$ as $|\psi\rangle = \sum_k C_k |a_k\rangle$ [27] where $C_k$ is a complex coefficient and $\sum |C_k|^2 = 1$. Based on the 3[rd] postulate, $\hat{A}$ is Hermitian operator associated to the observable $a$. We recall, $\hat{A}|a_k\rangle = a_k|a_k\rangle$, and whenever the measurement is performed, the system is thrown into one of the Eigen state of observable $\hat{A}$ (wave function collapse).

"*A measurement always causes the system to jump into an eigenstate of the dynamical variable that is being measured.*" P.A.M. Dirac [13].

In order to address the notion of measurement, we highlight that it is an interaction between a quantum system (measured) and a classical system (measuring).

In the ideal measurement theory developed by Von Neumann [23], he emphasized that quantum mechanics is a universally applicable theory. So, any physical system is basically a quantum mechanical system. To have a consistent theory of measurement, we must, therefore treat the apparatus $A$, quantum mechanically [32]. So, we consider a quantum system $Q$, represented by basis vectors $\{|q_k\rangle\}$ in a Hilbert space $H_q$, interacts with a measurement apparatus $A$, described by basis vectors $\{|a_k\rangle\}$ spanning a Hilbert space $H_A$. In this case, $|a_k\rangle$ are assumed to correspond to macroscopic "pointer". So, the state space of composite system–apparatus is given by tensor product of individual Hilbert spaces $H_q \otimes H_A$.

We suppose, the quantum system is described by $|\Psi\rangle$. In this case, it is in a superposition state $|\psi\rangle = \sum_k C_k |q_k\rangle$ and $A$ is in the initial state $|a_i\rangle$ before measurement. By the way, due to the linearity of the Schrödinger equation, the total system $QA$, assumed to be represented by the Hilbert product space $H_q \otimes H_A$ and the composite system-apparatus could be described as follows

$$|\psi_{QA}\rangle = \left(\sum_k C_k |q_k\rangle\right) \otimes |a_i\rangle \tag{2.1}$$

which after an infinitesimal time evolution $t$, we obtain

$$|\psi_{QA}\rangle = \sum_k C_k |q_k\rangle |a_k\rangle \tag{2.2}$$

As the considered system consist of a sum over more than one states, the system is in a statistical mixture and it is in a superposition of the basis states $|q_k\rangle \otimes |a_k\rangle$. The density matrix (quantum equivalent of a probability density) is then given by [30]

$$\rho = |\psi_{QA}\rangle\langle\psi_{QA}| = \sum_{k,k'} p_k |q_k\rangle|a_k\rangle \otimes \langle q_{k'}|\langle a_{k'}| \tag{2.3}$$

where $p_k = |C_k|^2$.

Equation (2.3) shows correlation between state of the pointer basis and corresponding relative state of the system which are preserved in the final mixed state density matrix [29].

Von Neumann even considered the possibility of collapse precipitated by a conscious observer [32]. In section 3, we survey Von Neumann Quantum measurement theory reformulation by Zurek leading to decoherence.

## 3. Zurek Reformulation and Decoherence

Based on Von Neumann quantum measurement theory, an additional quantum apparatus $A'$, coupled to the original $A$, cannot be of any help in resolving the difficulties of the measurement problem. While Von Neumann has considered the quantum system, $Q$ and apparatus $A$, as isolated from the rest of universe, the discussions have paid a lot of attention to the immersion of the apparatus in their environments [29, 31- 40].

Zurek has shown that the interaction between the quantum apparatus $A$ and its environment $E$ may single out a preferred pointer basis of the apparatus [29]. To reach to the right conclusion, he has reconsidered the Von

Neumann model for ideal quantum mechanics measurement, but with the environment included. In the reformulation, one can consider the environment denoted by $\varepsilon$ and its state before the measurement as represented by initial state vector $|e_0\rangle$ in a Hilbert space $H_\varepsilon$ [41]. On the other hand, we know the state space of the composite object system – apparatus- environment is given by the tensor product of the Hilbert spaces, $H_q \otimes H_A \otimes H_\varepsilon$ (see section 2). In this case, one can consider the composite system -apparatus- environment as vector $|\psi_{QA\varepsilon}\rangle$, and based on the linearity of Schrödinger equation, the time evolution of the system reads

$$|\psi_{QA\varepsilon}\rangle = \left(\sum_k C_k |q_k\rangle\right) \otimes |a_i\rangle \otimes |e_0\rangle \tag{3.1}$$

After two stage infinitesimal time evolution, this gives

$$|\psi_{QA\varepsilon}\rangle = \left(\sum_k C_k |q_k\rangle|a_k\rangle\right) \otimes |e_0\rangle \xrightarrow{t} |\psi_{QA\varepsilon}\rangle = \sum_k C_k |q_k\rangle|a_k\rangle|e_k\rangle \tag{3.2}$$

So, we obtain the density matrix as follows

$$\rho = |\psi_{QA\varepsilon}\rangle\langle\psi_{QA\varepsilon}| = \sum_{k,k'} C_k C_{k'}^* |q_k\rangle|a_k\rangle|e_k\rangle \otimes \langle q_{k'}|\langle a_{k'}|\langle e_{k'}| \tag{3.3}$$

As, we cannot assume the basis vectors $|e_k\rangle$ of the environment are necessarily mutually orthogonal, one can reformulate (3.3) as (for detail see Ref. [42])

$$\rho_{QA\varepsilon} = \sum_k |C_k|^2 |q_k\rangle|a_k\rangle|e_k\rangle \otimes \langle e_k|\langle a_k|\langle q_k|. \tag{3.4}$$

Based on (3.4), Zurek has described, the interaction between $Q$, $A$ and $\varepsilon$ is a process in two stages. The first one consists of a unitary time evolution of the state vector of universe, leading to a pure correlation between state vectors $Q$ and $A$. The second consist of a unitary time evolution of the state of universe transforming the pure correlation between state vectors $Q$ and state vectors of $A$ [57]. The first stage is called "Decoherence" and the second one known as "super selection "by the pioneers.

Another more recent aspect of the Zurek's theory term "invariance" was proposed in [32, 35] and further developed in [35]. In this case, Zurek used invariance to explain the emergence of probabilities in quantum mechanics and derive Born's rule based on certain assumptions. Finally, we would like to emphasize that Zurek measurement theory arises from a direct application of the quantum mechanical formalism to a description of the interaction of physical system with the environment.

Although, the Zurek measurement has showed some success during the years before leading to decoherence, the core difficulty associated with consistent histories, needs to clarify the emergence of the classical world of our experience from the underlying quantum nature [42].

## 4. Fuzzy Logic and Fuzzy Set Theory

The Greek philosopher Aristotle (348 B.C.E- 322 B.C.E) founded a system of logic based on only two types of proposition, true and false [43-45]. The Aristotle binary logic always leads us to draw the line between opposites, A or not A.

The English mathematician George Boole (1815-1864) sought to give symbolic form to Aristotle logic [46]. He codified several rules of two possible values mathematics, 1 or 0. The proposed system by Boole became known as Boolean algebra. It captures essential properties of both set operations and logic operations. By the way, the set theory was founded by George Cantor by a single famous paper in 1874 [47, 48]. The set theory begins with a fundamental theory in the notion of membership. In fact, sets have members which also called elements [56]. In this case, we suppose A as a subset of U, if $A \neq U$, then there are two categories of elements which are not member of A. In case, one can define a function of subset A of set $U$ as $f_A : U \to \{0,1\}$ which is called indicator function as

$$f_A(x) = \begin{cases} 1 & \text{if } x \in A \\ 0 & \text{if } x \notin A \end{cases} \tag{4.1}$$

Based on the set theory, the interpretation of universe is just black and white but here, an important question raises as: Is this interpretation consistent with reality?

In order to address the above question, we consider a set of students. Everybody who is a student is a member of this set otherwise is not. Now, we consider the set of all employees of a company who are tall. What can we say about the membership of a specific person? To find the right answer, one need to look into Fuzzy sets theory and Fuzzy logic. The idea of Fuzzy sets was introduced by Lotfi A. Zadeh in 1965 [49] and the Fuzzy logic was developed later by Zadeh in 1975 [50].

A Fuzzy set was defined [58,59,60] as a class of objects with a continuum of grades of membership [51]. It is a pair $(N,\mu)$ where $N$ is a set and $\mu: N \to [0,1]$. For each $x \in N$, the value $\mu(x)$ is called the grad of membership of $x$ in $(N,\mu)$. Let $x \in N$; then $x$ is not included in the fuzzy set $(N,\mu)$ if $\mu(x)=0$ and $x$ is fully included if $\mu(x)=1$. In case, if $0 \prec \mu(x) \prec 1$ then $x$ is a fuzzy member [53-55]. The set $\{x \in N | \mu(x) \succ 0\}$ known as the support of $(N,\mu)$ and the set $\{x \in N | \mu(x) = 1\}$ is called its Kernel. In this regards, the function $\mu(x)$ is membership function of the fuzzy set. Based on the Fuzzy set theory, the universe is not black and white but is gray; a continues range between black and white and everything is a matter of degree [52]. The above description highlight that the gray universe is consistent with reality.

## 5. Fuzzy Quantum Measurement

We recall that Von Neumann quantum measurement theory has developed based on two parts, quantum system, and apparatus and the measuring apparatus obeys the rules of quantum mechanics similar to the measured system. In this way Zurek has developed his new theory of measurement considering the environment as a new interacting part. So, the environment was added to the measurement theory. Zurek had assumed; these three portions, quantum system, apparatus, and environment obeying the rules of quantum mechanics but not including of the observers. He highlighted, the observer typically does not interact with systems of interest themselves, but with their surrounding environments.

*So, is the system identified by observer as a part of environment and apparatus characterized by possible outcome values of their quantum degree of freedom, or by possible outcome values of bulk degrees of freedom such as macroscopic size?*

Here, we consider four identical particles, *A, B, C & D* as one dimensional interacting chain (see Fig. 1).

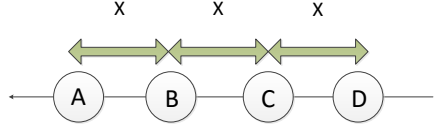

Fig.1. Schematic interacting particles

Three interacting systems by the particles $A \rightleftharpoons B$, $A \rightleftharpoons C$ and $A \rightleftharpoons D$ could be considered. We quantify the interaction by parameter $F$. In this case, one can find

$$F_{\overrightarrow{AB}} \succ F_{\overrightarrow{AC}} \succ F_{\overrightarrow{AD}}$$
$$F_{\overrightarrow{AC}} = \frac{1}{2} F_{\overrightarrow{AB}}, \quad F_{\overrightarrow{AD}} = \frac{1}{3} F_{\overrightarrow{AB}} \qquad (5.1)$$

Now, we consider $F$ as a fuzzy set of the interaction parameters $F$. If the value of $F_{\overrightarrow{AB}}$'s membership function, $m_{F_{\overrightarrow{AB}}}$, could be set to 1 then based on (5.1), the amount of membership functions could be as follows

$$m_{F_{\overrightarrow{AC}}} = \frac{1}{2} = 50\% \quad \& \quad m_{F_{\overrightarrow{AD}}} = \frac{1}{2} = 30\%. \qquad (5.2)$$

Let us go back to the Zurek theory. In this case, same as the above example, obviously, the percentage of the participation of interaction between quantum system-environment, quantum system-apparatus and apparatus-environment are not equal. We consider, a measurement set up as figure 2.

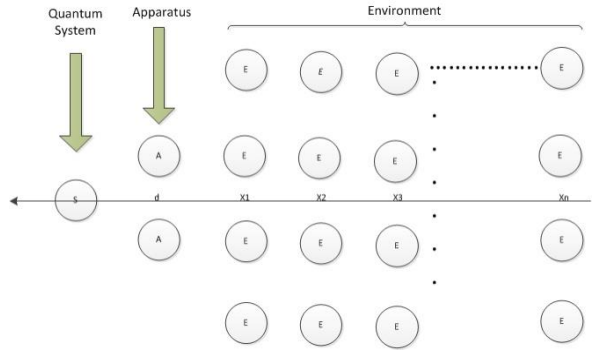

Fig.2. Schematic figure of measuring system

It is evident that interaction between the apparatus particles and the environment particle which is localized in $x_1$, quantitatively, is not equal to the interaction between the apparatus and the particle localized in $x_2$. We suppose, the quantity of interaction between the apparatus and the particle located in $x_n$ is indicated by $F_{dX_n}$ and the interaction is dependent on the particles distance directly.

So, one cannot consider the apparatus and the environment as two parts which participate completely in the interaction with the quantum system. We develop our new idea based on the fuzzy logic and fuzzy set theory which is proposed in section 4 by the following argument.

We consider the universe as a fuzzy reference set which is indicated by $U$ [52]. We suppose; there is a fuzzy subset of $U$, and the fuzzy set of the quantum world indicated by $Q$. So, $Q \subset U$ and the apparatus and environment are the members of $Q$, since $Q$ is considered as a fuzzy subset. In this case, the membership function for apparatus and environment should be considered. We know that the universe is a physical system which could be characterized by a Hilbert space $H$. In order to develop of mathematical image of the idea, one may consider fuzzy Hilbert space or in the other words fuzzy inner product space.

**Definition 1** Let $V$ a fuzzy linear space. If, for an arbitrary pair of elements u and v, there is a number $(u,v)$ of [0, 1] such that satisfies:

1) $(u,v) = (v,u)$
2) $(ku,v) = k(v,u) \quad k \in [0,1]$
3) $(u+v,w) = (u,w) + (v,w) \quad w \in V$
4) $(u,u) = 0 \quad \text{if} \quad u = \theta$

Then $V$ is called a fuzzy inner product space, and $(u,v)$ is called the fuzzy inner product of u and v. In fuzzy inner product space, the following hold

1) $(ku,hv) = kh(u,v) \quad k,h \in [0,1]$
2) $(u, kv+hw) = k(v,u) + h(u,w) \quad k,h \in [0,1]$
3) If $u$ or $v$ is 0 then $(u,v) = 0$
4) $\left( \sum_{i=1}^{m} k_i u_i, \sum_{j=1}^{n} h_j V_j \right) = \sum_{i=1}^{m} \sum_{j=1}^{n} k_i h_j (u_i, v_j) \quad k_i, h_j \in [0,1]$

In the finite spanning inner product space $V$, let $\{e_1,...,e_n\}$ be a basis of fuzzy inner product space V, for arbitrary $u \in V$, $v \in V$ if $u = x_1 e_1 + ... + x_n e_n$, $v = y_1 e_1 + ... + y_n e_n$ then $(u,v) = \left( \sum_{i=1}^{n} x_i e_i, \sum_{j=1}^{n} y_j e_j \right) = \sum_{i=1}^{n} \sum_{j=1}^{n} (e_i, e_j) x_i y_j$. Let $a_{ij} = (e_i, e_j)$ and $A = (a_{ij})_{m \times n}$ then

$$XAY^T = (x_1...x_n) \begin{bmatrix} a_{11} & \cdots & a_{1n} \\ \vdots & & \vdots \\ a_{n1} & \cdots & a_{nn} \end{bmatrix} \begin{pmatrix} y_1 \\ \vdots \\ y_n \end{pmatrix}$$

$$= \sum_{i=1}^{n} \sum_{j=1}^{n} a_{ij} x_i y_j = (u,v) \quad (5.3)$$

That is $(u,v) = XAY^T$, where $X = (x_1,...,x_n)$ and $Y = (y_1,...,y_n)$.

**Definition 2** Let $\{e_1,...,e_n\}$ be a basis of a fuzzy inner product space V and $a_{ij} = (e_i, e_j)$ then $A = (a_{ij})$ is called a metric matrix of V under the basis $\{e_1,...,e_n\}$.

**Theorem 1** Let two bases $\{e_1,...,e_n\}$ and $\{v_1,...,v_n\}$ of fuzzy inner product space V and A is a metric matrix of V under the basis $\{e_1,...,e_n\}$ and B is a metric matrix of V under basis $\{v_1,...,v_n\}$ if C is a transition matrix from $\{e_1,...,e_n\}$ to $\{v_1,...,v_n\}$ that is $(v_1,...,v_n) = (e_1,...,e_n)C$ then $B = (b_{ij})_{n \times n} = ((v_i, v_j))_{n \times n} = C^T AC$.

**Definition 3** For two fuzzy matrices A and B if there is a fuzzy matrix C such that $B = C^T AC$ then $B$ and $A$ is called similar.

So, we could develop the fuzzy inner product space and its characteristics. Let $\{x_k\}_{k=1}^{\infty}$ be a sequence of elements of $H$. Based on the fuzzy algebra and the proposed definitions, if $(H, \langle u,v \rangle)$ is an ordinary inner product space then $(H, \langle u,v \rangle k)$ is fuzzy inner product space where $H$ is a vector space over field F. So, fuzzy Hilbert space could be considered as a subspace of ordinary Hilbert space, we consider a fuzzy operator $M$ which denote the weight of interaction for each part of environment, apparatus and quantum system with each other. Based on the quantum mechanics algebra, the following equation could be considered in fuzzy Hilbert space:

$$M|m'\rangle = m'|m'\rangle \tag{5.4}$$

In this regards, one can find the weight of interaction between the environment; apparatus and the system by operating (5.4) on the system. As the weights of the participation are fixed for each measurement set up so the sequence of operation between the operator $M$ and the observable operator like Hamiltonian is not important i.e. the fuzzy operator $M$ and any other operators denoted by observables are commutative.

**Definition 4** Let $V_{m \times n}$ denote set of all $m \times n$ fuzzy matrices [61] over the fuzzy algebra $F = [0,1]$. The operations $(+,.)$ are defined on $V_{m \times n}$ as follows:

i. For any two elements $A(a_{ij})$ and $B(b_{ij}) \in V_{m \times n}$ define $A + B = (\sup\{a_{ij}, b_{ij}\}) = \vee_{i,j}(a_{ij}, b_{ij})$ where each $a_{ij}, b_{ij} \in F$

ii. For any element $A(a_{ij}) \in V_{m \times n}$ and a scalar $k \in F$ define $kA = (\inf\{k, a_{ij}\}) = \wedge_{i,j}(k, a_{ij})$

The above definition indicates that the operation on fuzzy Hilbert space can be done by matrices.

**Definition 5** Let $V_{m \times n}$ be a fuzzy normed linear space over $F$. A mapping $T$ from $V_{m \times n}$ into itself is called a fuzzy linear transformation (operator) if for any $A, B \in V_{m \times n}$ and $k \in F$

i. $T(A+B) = T(A) + T(B)$

ii. $T(kA) = kT(A)$

**Remark:** Let $V_{m \times n}$ be a fuzzy normed linear space over $F$. Then $L(V_{m \times n})$, the set of all linear transformation from $V_{m \times n}$ to itself is a vector space and fuzzy addition and fuzzy multiplication are defined as

i. $(T_1 + T_2)A = T_1(A) + T_2(A)$

ii. $(\alpha T_1)A = \alpha T_1(A)$

For all $T_1, T_2 \in L(V_{m \times n})$, $A \in V_{m \times n}$ and $\alpha \in F$. In this case, there are some important properties for them as follows

i. $(\alpha \beta)T_1 = \alpha(\beta T_1)$

ii. $(\alpha + \beta)T_1 = \alpha T_1 + \beta T_1$

iii. $\alpha(T_1 + T_2) = \alpha T_1 + \alpha T_1$

iv. $T_1 T_2(x) = T_1(T_2(x))$

v. $T_1 + 0 = T_1$

**Definition 6** Let $V_{m \times n}$ be the fuzzy normed linear space. Let $T$ be the fuzzy linear transformation from $V_{m \times n}$ to itself. If there exist an operator $T^*$ such that $\langle T(A), B \rangle = \langle A, T^*(B) \rangle$ for all $A, B \in V_{m \times n}$ then the operator $T^*$ is called fuzzy adjoint of $T$. Now, consider $A'$ as observable operator then

$$[A', M] = 0 \tag{5.5}$$

where the observable $A'$ and $M$ are compatible. They have simultaneous eigenket [13]. We consider the above sequence as an ensemble of fuzzy set $U$. Let $|A_0\rangle$ be the initial state of apparatus and at the same time $|\varepsilon_0\rangle$, the initial state of the environment. Then, one can consider the projection of $|A_0\rangle$ which is an element of the Hilbert space as an ensemble of the universe fuzzy set. So, it takes the form $|A_0\rangle = \sum_{k=1}^{\infty} a_k |x_k\rangle$.

In this case, $M_n$ could be a Hilbert space considered as a linear manifold spanned by $x_1, x_2, ..., x_n$: $M_n = span(\{x_k\}_{k=1}^{n})$ and it is an ensemble of the quantum world fuzzy set denoted by $Q$. Based on the above recipe, first we can operate $M$ on the State ket of the measurement set up.

$$M|A_0\rangle = M|m'\rangle = \sum_{k=1}^{n} m^{A_k} |m'\rangle \tag{5.6}$$

and one can write the state ket of the apparatus as

$$|A_0\rangle = \sum_{k=1}^{n} a_k m^{A_k} |A_k\rangle \tag{5.7}$$

then, the measurement set up Eigen ket reads

$$|\psi_{SA\varepsilon}\rangle = |\psi_S\rangle \otimes |A_0\rangle \otimes |\varepsilon_0\rangle \tag{5.8}$$

and after operating the fuzzy operator, we obtain

$$|\psi_{SA\varepsilon}\rangle = |\psi_S\rangle \otimes |A_0\rangle \otimes |\varepsilon_0\rangle = \sum_{k=1}^{n} C_k m^{S_k} |S_k\rangle m^{A_k} |A_k\rangle m^{A_k} |e_k\rangle \tag{5.9}$$

So,

$$|\psi_{SA\varepsilon}\rangle = \sum_{k=1}^{\infty} \prod_{j=S,A,E} m^{j_k} C_k |S_k\rangle |A_k\rangle |e_k\rangle \tag{5.10}$$

In the above equation, $\prod_{j=S,A,E} m_Q^{j_k}$ is Fuzzy Quantum Measurement Coefficients (FQMC). In order to find the exact value of measurement for quantum system, FQMC should be calculated. To explore the physical consequences of the non uniqueness of the basis chosen to represent an environment; apparatus correlated with a quantum system, we consider a simple Gedanken experiment for better understanding. In this case, we divide the experiment into three steps.

First, we consider two – state quantum apparatus used to "measure "the other two-state quantum system. Stern-Gerlach magnets can be arranged to split the spin $-\frac{1}{2}$ beam. We supplement it by a bi-stable atom acting as a quantum apparatus same as what has been employed by Scully, Shea, McCullen and Zurek [62, 29] who have used a bi-stable atom as a microscopic model of "Wigner's friend" [63]. We suppose the Stern- Gerlach magnets as apparatus, and quantum system fully correlated. so, their correlation percentage is 100%. If a spin enters SG in an initially pure Eigen-state, $|\Psi\rangle = \frac{(|\uparrow\rangle + |\downarrow\rangle)}{\sqrt{2}}$ then the apparatus- quantum system initial Eigen-state can be represented by

$$|\Psi_{QA}\rangle = |\Psi\rangle \otimes |A\rangle \tag{5.11}$$

When the spin- atom (quantum apparatus) interaction begins for affecting the 100% correlation into initial state one might use the following Fuzzy Quantum Measurement operators

$$M_Q^S = \begin{bmatrix} 1 & 0 \\ 0 & 1 \end{bmatrix} \quad \text{and} \quad M_Q^A = \begin{bmatrix} 1 & 0 \\ 0 & 1 \end{bmatrix} \tag{5.12}$$

So, the initial state remains unchanged.

Here, we add another SG into our setup as second step. We set the magnetic fields of the first apparatus which affected into the quantum system along $z$ axis of the coordinate system. So, if the magnetic fields of the second one, is set along $z'$ axis of the other coordinate system which has the same $x$ axis with the first one, then we can define $\theta$ as angle between $z$ and $z'$. In case, one can set, $\theta = 45°$. It is obvious, the magnetic field of second apparatus is in correlation with quantum system not as complete as the first one. It is straightforward to calculate the correlation percentage which could be 50% due to $\cos\theta = \frac{1}{2}$.

In order to affect 50% correlation into initial state, we use the FQM operator as follows

$$M_Q^{A'} = \begin{bmatrix} \frac{1}{2} & 0 \\ 0 & \frac{1}{2} \end{bmatrix} \quad (5.13)$$

More over; the environment correlation with our setup depends on the environment components positions.

Here, considering three particles as the environment components turn us into the third step of Gedanken experiment. These three particles; apparatus and quantum system are isolated. If the environment particles correlate with apparatus and quantum system based on short range interaction potential, then, the correlation could be considered as a function of $\vec{r}$ which is the position of the environment particles. In this case; one can affect the weight of correlation between the environment, apparatus and quantum system by the following FQM operator.

$$M_Q^{\varepsilon A} = \begin{bmatrix} f(\vec{r}_Q - \vec{r}_A) & 0 \\ 0 & f(\vec{r}_Q - \vec{r}_A) \end{bmatrix} \quad (5.14)$$

where the $\vec{r}_Q$ is the position of apparatus – system and $\vec{r}_A$ is the position of the particle indicated by A. The dimension of the matrix which is representative of the operator, depends on the number of the environment particles and dimensions of the fuzzy Hilbert space.

## 6. Conclusion & Discussion

The Von Neumann quantum measurement theory was developed based on the important assumption which state the quantum system and apparatus obey the quantum mechanics rules both. This assumption was considered as a base by Zurek in his reformulation for quantum system, apparatus and the environment which was added by him as new interacting part of measurement set up. In this article we have changed this main assumption into a new one. Although the assumption which state the universe obeying the quantum mechanics rules seems correct, but we believe in a measurement set up, just the interacting part of the universe with measured quantum system could be considered as part of the quantum world. As a conclusion, we have proposed new correction coefficients called FQMC. Based on the Von Neumann and Zurek, the measurement is a process in two stages. The first stage consists in a unitary time evolution of the state vectors of the environment leading to a pure correlation between state vectors of quantum system and apparatus. The second one, consist in a unitary time evolution of the state of the universe, transforming the pure correlation between state vectors of quantum system and state vectors of apparatus into a pure correlation between state vectors of quantum system and state vectors of apparatus but it is a continues process occurring within a finite time interval [57]. When the quantum system and apparatus combine and left in a pure state that represents a linear superposition of system-pointer states, we cannot realize the system actually is in which states after collapse. In this case, based on Fuzzy Quantum Measurement, the actual state would be nominated by whole combined system-apparatus-environment based on the interaction between the components of each part. The second difficulty associated with quantum measurement is known as the preferred basis problem. In general, for some choice of apparatus states, the corresponding new system states are not mutually orthogonal. So that the observable associated with these states is not Hermitian, and is usually not desired. The Fuzzy Quantum Measurement could be able to clear this problem also due to effect of FQM operator. By affecting these operators, the base also could be nominated exactly related to the number of interacting particles which are engaged in the setup. In fact, by FQMC correction, the base state problem was removed due to affect a particular consequence recognizing by FQMC on the correlation between the universes, apparatus and the quantum system state vectors. The sequence of correlation is based on the value of FQMC. In this case, If $m^A > m^\varepsilon$, the first step of correlation is between the quantum system state vectors and apparatus and so on. For instance, doing measurement on a located atom in a solid network, the surrounding phonons and other neighbor atoms would act more likely as an environment. In this case $m^A < m^\varepsilon$, and the first stage correlation could be between quantum system state vectors and environment state vectors.

### Acknowledgment

SR would like to acknowledge grant IP-2014-054 support by UKM.

### References

[1] Born. M. and Wolf. E. 1999. Principle of optics. Cambridge University press.

[2] Planck. M. 1900. Entropy and temperature of radiant heat. Annalen der physic. vol.1. no4: 719-37

[3] Planck. M. 1901. On the law of the distribution of energy in the normal spectrum. Annalen der physic. vol.4: p.553ff

[4] Einstein. A. 1905. Über einen die Erzeugung und verwand lung des litchtes betretfenden heuristischen Gesichtspunkt (on a Heuristic Viewpoint concerning the production and transformation of light). Annalen der physic. 17 (6):132-148

[5] Planck. M. 1914. the theory of heat radiation. (second edition) translated by Masius. M. pages 22,26,42,43 Blakiston's son &co, Philadelphia

[6] Albrecht. F. and Einstein. A. 1997. A Biography, Trans. Ewald Osersi, Viking

[7] Hanle. P. A. 1977. Erwin Schrödinger's Reaction to Louis de broglie's thesis on the quantum theory. Isis 68(4): 606-609

[8] Bacciagalup. G. Valentini. A. 2009. Quantum theory at the crossroads: reconsidering the 1927 Solvay conference. pp. 9184. Cambridge university press, Cambridge

[9] Jammer. M. 1966. The conceptual development of quantum mechanics. MC Grow–Hill, New York

[10] Jammer. M. 1974. The philosophy of quantum mechanics: the interpretations of quantum mechanics in historical perspective. Wiley, New York

[11] Griffith. D. J. 2004. Introduction to Quantum Mechanics. (2nd edition) Pearson Prentice Hall

[12] Cohen-Tannoudji. C. and Bernard Diu. L. F. 1997. Quantum mechanics (2 Vol. set). (2nd edition) Hermann and John Wiley & sons Inc

[13] Sakurai. J. J. and Napolitano. J. J. 2010. Modern Quantum Mechanics. (2nd edition) Addison-Wesley

[14] Merzbacher. E. 1969. Quantum Mechanics. John Wiley & sons Inc

[15] Schiff. J. I. 1968. Quantum Mechanics. McGraw-Hill Education

[16] Hecht. K. T. 2008. Quantum Mechanics. springer (India) Pvt. Ltd.

[17] Liboff. L. 1991. Introductory quantum mechanics. (4th edition) Addison-Wesley

[18] Tegmark. M. and Wheeler. A. 1999. 100 years of quantum mysteries. pp.68-75, scientific American

[19] Griffiths. R. B. and Omnès. R. 1999. Consistent Histories and Quantum Measurements. physics today. pp.26-31

[20] Goldstein. S. 1998. Quantum Theory without observers. pp.42-46

[21] Yam. P. 1997. Bringing Schrödinger's cat to life. pp.124-129. scientific American.

[22] Heisenberg. W. 1949. The physical principle of quantum theory. Chicago university: 1930. the physical principles of the quantum theory. New York: Dover

[23] Von Neumann. J. 1955. Mathematical foundations of Quantum mechanics. springer: 1932. Mathematical foundations of quantum mechanics. Princeton university press

[24] Klyshko. D. N. 1998. Reduction of wave function. laser physics. vol.8. no. 2, pp.363-389

[25] Brune. M. Hagley. E. Dreyer. J. Maitre. X. Maali. A. Wanderlich. C. Raimond. J. M. and Haroche. S. 1996. Observing the progressive decoherence of the meter in a quantum measurement. phys. Rev. Lett. 77, 4887-4890

[26] Dorobantu. V. 2005. *The postulates of quantum mechanics, quantum computability*. vol.1. politechnica university press

[27] Debanth. L. and Mikusinski. P. 2005. *Hilbert spaces with applications*. (third edition) Elsevier Academic press

[28] Busch. P. Johannes Lati. P. and Mittelstaedt. P. 1996. *The quantum theory of measurement*. springer

[29] Zurek. W. H. 1981. *pointer basis of quantum apparatus: Into what mixture does the wave packet collapse?*. Vol. 24. no 6. physical Review D

[30] Giulini. D. Joos. E. Kiefer. C. Kupsch. J. Stamatescu. I. O. and Zeh. H. D. 1996. *Decoherence and the appearance of a classical world in quantum theory*. springer

[31] Paz. J. P. Zurek. W. H. 2001. Environment-induced decoherence and the transition from quantum to classical Course 8 (pp. 533-614) of *Les Houches Lectures Session LXXII: Coherent Atomic Matter Waves*., Kaiser. R. Westbrook. C. David. F. eds. Springer, Berlin

[32] Zurek. W. H. 2003. Decoherence, Einselection and the quantum origins of the classical. Rev. Mod. Phys. 75, 715-765

[33] Zurek. W. H. 2000. Ann. der physic, (Leipzig) 9, 855

[34] Ollivier. H. Poulin. D. Zurek. W. H. 2004. Objective properties from subjective quantum states: Environment as a witness. *Phys. Rev. Lett.* **93**, 220401

[35] Zurek. W. H. 2003. Environment-Assisted Invariance, Ignorance, and Probabilities in Quantum physics. Phys. Rev. Lett. 90, 120404

[36] Zurek. W. H. 1993. Progr. Theor. Phys. 89, 281-312

[37] Zurek. W. H. 1998. Phil. Trans. Roy. Soc. Land. A 356. 1793

[38] Zurek. W. H. 1982. Phys. Rev., D26, 1862-1880

[39] Zurek. W. H. 1991. Physics Today 44, 36-44

[40] Zurek. W. H. 1994. Decoherence and the Essential interpretation of Quantum Theory or: No Information without Representation. pp. 341-350 of from statistical physics to statistical Inference and Back., Grassberger. P. Nadal. J. P., eds. Plenum. Dordercht

[41] Schlosshauer. M. 2005. The 'self-induced decoherence' approach: Strong limitations on its validity in a simple spin bath model and on its general physical relevance. *Phys. Rev. A* **72**, 012109

[42] Schlosshauer. M. 2004. Decoherence, the measurement problem and interpretation of quantum mechanics. Rev. Mod. Phys. Vol. 76. no 4.

[43] Prior. A. N. 1962. Formal Logic. Q.U.P. Oxford

[44] Thompson. M. 1953. On Aristotle's square f opposition. philosophical Review 62. 251-265

[45] Slater. B. H. 2009. Hilbert's epsilon calculus and its successor's. pages 385-448 in: Gabbay. D. and Woods. J. (eds) Handbook of The History of logic. Vol. 5. Elsevier Science. Burlington MA

[46] Boole. G. 1854. An investigation of the laws of thought on which are founded the mathematical theories of logic and probabilities. Walton and Maberly. London

[47] Cantor. G. 1854. Ueber eine eigenschaft des inbegriffes aller reellen algebraischen Zahlen. J. Reine Angew. Math. 77: 258-262

[48] Johnson. P. A 1972. History of set theory. Prindle. weber & Schmidt

[49] Gabbay Dov. M. Woods, J. 2007. Fuzz-Set based logics-An History-oriented presentation of their main Developments. Handbook of the history of logic. Elsevier

[50] Zadeh. L. 1975. Fuzzy Logic and approximate reasoning. Syntheses. 30. 407-428

[51] Zadeh. L. 1965. Fuzzy set. information and control. Vol. 8. 338-353

[52] Kosko. B. 1993. Fuzzy thinking: the new science of fuzzy logic. Hyperion. New York

[53] Dubois. D. and Prade. H. 1980. Fuzzy set and systems: theory and applications. Academic press. New York

[54] Zimmermann. H. J. 1993. Fuzzy set theory and its applications. Academic publishers. Kluwer

[55] Zadeh. L. Gklir. G. J. and Bo. Y. 1996. Fuzzy sets, Fuzzy logic, and Fuzzy systems: selected papers. world scientific publishing co Pte. Ltd.

[56] Hajnal. A. Hamburger. P. 1999. Set theory. Cambridge University press

[57] Matehkolaee. M. J. 2011. Understanding the measurement theory in quantum mechanics. lat. Am. J. phys. Educ. Vol. 5. No 3.

[58] Borkotokey. S. 2010. Advanced topics in fuzzy Algebra: recent Developments. VDM verlag Dr. Muller.

[59] Mordeson. J. N. and Malik. D. S. 1998. Fuzzy commutative Algebra (pure Mathematics). World Scientific

[60] Vasantha Kardasamy. W. B. Smarandache. F. and Ilanthenral. K. 2009. Special set linear Algebra and special set fuzzy linear Algebra. Editura CuArt and Authors

[61] Vasantha Kardasamy. W. B. Smarandache. F. and Ilanthenral. K. 2007. Elementary fuzzy matrix theory and fuzzy models for social scientists. Automaton, Kardasamy, Smarandache. Ilanthenral.

[62] Scully. M. Shea. R. and McCullen. J. D. 1978. Phys. Rep. 43C. 485.

[63] Wigner. E. P. 1967. In The Scientist Speculates. edited by I. J. Good (Heinemann. London. (1961), Basic Books. p. 284. New York (1962), reprinted in Wigner. E. P.: Symmerties and Reflections. p. 171 Indiana University Press. Bloomington and London